\documentstyle[preprint,aps]{revtex}
\begin{document}
\tighten
\title{ Threshold scattering of the $\eta$-meson off light nuclei}
\author{
S. A. Rakityansky\footnote{Permanent address:
Joint Institute  for Nuclear Research, Dubna, 141980, Russia},
S. A. Sofianos, and
W. Sandhas\footnote{Permanent address:
 Physikalisches Institut, Universit\H{a}t Bonn, D-53115 Bonn, Germany}}
\address{ Physics Department, University of South Africa,
 P.O.Box 392,Pretoria 0001, South Africa}
\author{V. B. Belyaev}
\address{ Joint Institute  for Nuclear Research,Dubna, 141980,
Russia}

\date{\today}
\maketitle

\begin{abstract}
The scattering lengths of $\eta$-meson collisions with light nuclei
$^2H$,\,$^3H$,\, $^3He$, and $^4He$ are
calculated on the basis of few-body equations
in coherent approximation.  It is
found that the $\eta$-nucleus scattering length depends
 strongly on the number of nucleons and the potential-range parameter.
By taking into account the off-shell behavior  of the $\eta N$amplitude,
the $\eta\alpha$ scattering length increases considerably.
\end{abstract}
\vspace{.5cm}

In $\eta$-nucleus collisions at threshold energies,
a large Final State
Interaction (FSI) is expected due to the
$N^*(1535)$ \,$ S_{11}$-resonance. This resonance, which is strongly
coupled
to the $\eta N$ and $\pi N$ channels, lies only
$\sim 50$ MeV above the $\eta N$ threshold
and has a very broad
width of
$\Gamma \approx 150 $ MeV \cite{PDGr}.
The $\eta$-nucleus dynamics, thus,
is of interest  from the point
 of view of both few-body and meson-nucleon
physics.\\

In the energy region covering the $S_{11}$-resonance, the $\pi N$
and $\eta N$ scattering processes are to be
considered as a coupled-channel problem
\cite{Bhal,Benn,Mcle}. The first of these channels has a long record
of theoretical and experimental investigations, while the second
one is understood only in a rudimentary way.
The $\pi N$ and $\eta N$ channels are
connected to each other primarily
via the $N^*(1535)$ resonance. In fact, the
$\eta NN$ coupling constant was shown to be
negligible \cite{Grei,Alva,Tiat} as compared to the one of the $\eta NN^*$
vertex.  This latter vertex constant, being
hence crucial for determining the
$\eta N$ interaction, is
known only with a large uncertainty,
so that the $\eta N$ scattering length
inferred from it varies from (0.27 + i 0.22)fm \cite{Bhal}
to (0.55 + i 0.30)fm \cite{Wilk}. The  $\eta$N
near-threshold interaction, therefore,
remains  an interesting field of
investigation. Of particular relevance in this context
is the possibility of $\eta$-nucleus
bound states \cite{Haid,Chri}. Their
existence would provide us with an
excellent opportunity to answer
the above-mentioned questions in a
reliable manner.\\

Various experimental groups have studied the
near-threshold production of
$\eta$-mesons in photo-  and electro - nuclear processes
\cite{L21,L22}, or in
$\pi$-nucleus \cite{Peng}, $ N$-nucleus \cite{Berg,Chia},
and nucleus-nucleus \cite{Fras} collisions.
The expected strong $\eta$-nucleus
FSI is evident in these measurements and must not be ignored in
theoretical
treatments.\\

A simple way of taking into account the low-energy FSI effects in
$\eta$-production processes
 is described in Ref. \cite{Goldb}.  It consists in multiplying
the reaction amplitude by an energy-dependent factor which contains
the final-state scattering length as a parameter. This parameter, hence,
is essential in analyzing the increasing number
of such production data.  To the best of our knowledge, only the
 scattering-length calculations by Wilkin \cite{Wilk}
for
the $\eta$ collision on  $^3He$, $^4He$, and $^7Be$ exist
in the literature. There,
the lowest-order optical-potential method was used with a simple
$\eta$N amplitude having no energy or off-shell dependence.\\

In the present work we calculate
the $\eta\, ^2\!H$, $\eta\, ^3\!H$, $\eta\,
^3\!He$, and $\eta\, ^4\!He$ scattering
lengths on the basis of few-body
equations with separable $\eta$N amplitudes of the type suggested in
Refs.\cite{Bhal,Benn}. The equations employed are obtained within the
Finite-Rank Hamiltonian
Approximation (FRHA), proposed in Refs. \cite{Bel1,Bel2}
as an alternative to the multiple scattering or optical potential
theory. Developed originally for the treatment of pion-nucleus
scattering, this technique was  later on also applied to other questions of
few-body physics,
such as the $\Lambda$-nucleus bound state problem \cite{Bel3} or
the  $NN$-scattering  in a six quark model \cite{Bel2}.\\

The main idea of FRHA consists in separating
the motion of the projectile
and the internal motion of the nucleons inside the nucleus.  Correspondingly
the total Hamiltonian is split according to\\
\begin{equation}
             H=H_0+V+H_A \ ,
\end{equation}
where $H_0$ denotes the kinetic energy operator of the $\eta$-meson
relative to the center of mass of the nucleus,
$V=\sum_{i=1}^A V_i$ the sum of the individual
$\eta N$-interactions, and $H_A$ the Hamiltonian of the nucleus.
The scattering amplitude for the transition from the initial state
$|\vec{k},\psi_0>$ to the
final state $|\vec{k}',\psi_0>$, with $|\psi_0>$
being the ground state of the nucleus,
\begin{equation}
               H_A|\psi_0>={\cal E}_0|\psi_0>,
\end{equation}
is given by
\begin{equation}
           f(\vec{k}',\vec{k};z)=-\frac{\mu}{2\pi}\,
           <\vec{k}',\psi_0|T(z)|\vec{k},\psi_0>.
           \label{ampli}
\end{equation}
The $T$-operator in this amplitude satisfies the Lippmann-Schwinger equation

\begin{equation}
\label{bonn4}
            T(z)=V+V\frac{1}{z-H_0-H_A}T(z).
\end{equation}
Introducing an auxiliary operator $T^0$  by

\begin{equation}
\label{bonn5}
           T^0(z)=V+V\frac{1}{z-H_0}T^0(z),
\end{equation}
we obtain instead of Eq. (\ref{bonn4})
\begin{equation}
\label{bonn6}
           T(z)=T^0(z)+T^0(z)\frac{1}{z-H_0}H_A\frac{1}{z-H_0-H_A}T(z).
\end{equation}

Our main approximation consists in restricting the spectral
decomposition of $H_A$ to the ground state $|\psi_0>$,
\begin{equation}
H_A\approx {\cal E}_0|\psi_0><\psi_0|.
\end{equation}

\noindent
Physically this so-called coherent approximation \cite{Kerm},
which is widely
used in multiple scattering and optical potential approaches, means
that during the multiple scattering of the $\eta$-meson
the nucleus remains
unexcited. Equation (\ref{bonn6}) then reads
\begin{equation}
               T(z)=T^0(z)+{\cal E}_0T^0(z)|\psi_0>
		     \frac{1}{(z-H_0)(z-H_0-{\cal E}_0)}<\psi_0|T(z)\ ,
\label{tm2}
\end{equation}
and, after sandwiching it between the ground state and the
relative-momentum states, we obtain for the matrix elements $T({\vec k}',{\vec
k};z)
\enspace = \enspace <\vec{k}',\psi_0|T(z)|\vec{k},\psi_0>$
the integral equation

\begin{eqnarray}
\nonumber
       T(\vec{k}',\vec{k};z) &=&
       <\vec{k}',\psi_0|T^0(z)|\vec{k},\psi_0>\\
&+&
       {\cal E}_0\int \frac{d\vec{k}''}{(2\pi)^3}
       \frac{ <\vec{k}',\psi_0|T^0(z)|\vec{k}'',\psi_0>}{
        (z-\frac{{k''}^2}{2\mu})(z-
       {\cal E}_0-\frac{{k''}^2}{2\mu})}\,T(\vec{k}'',\vec{k};z).
\label{tm3}
\end{eqnarray}
Our problem thus is split into two steps. The first
consists in evaluating the auxiliary amplitude
$<\vec{k}',\psi_0|T^0(z)|\vec{k},\psi_0>$ which determines the inhomogeneity
and the kernel of Eq. (9); in the second step  this equation  is to be
solved.\\

It is easily seen that the operator
$T^0$ describes the scattering of the
$\eta$-meson off the nucleons which are fixed
in their position within the nucleus. This is due to the fact that
Eq. (\ref{bonn5}) does not contain any operator which acts on the
internal nuclear Jacobi coordinates
\{$\vec{r}$\}. All operators in Eq.
(\ref{bonn5}), hence, are diagonal in these variables, so that
its momentum representation reads
\begin{equation}
            T^0(\vec{k}',\vec{k};\vec{r};z)=V(\vec{k}',\vec{k};\vec{r})
                + \int \,\frac{d\vec{k}''}{(2\pi)^3}
                 \frac{V(\vec{k}',\vec{k}'';
                 \vec{r})}{z-\frac{k''^2}{2\mu}}
                 \,T^0(\vec{k}'';\vec{k};\vec{r};z).
\label{t0m}
\end{equation}

\noindent
It depends, in other words, only parametrically on the
coordinates $\{\vec{r}\}$. After solving this integral equation and
determining the ground state wave function  $\psi_0(\vec r)$  by any
appropriate bound-state method, the input
$<\vec{k}^\prime, \psi_0 | T^0(z)|\vec{k}, \psi_0>$ to Eq. (\ref{tm3})
is obtained via

\begin{equation}
<\vec{k}^\prime, \psi_0 | T^0(z) | \vec{k}, \psi _0 > = \int d\vec{r} |
\psi_0(\vec{r})|^2 T^0(\vec{k}^\prime,\vec{k};\vec{r};z).
\label{aver}
\end{equation}

Since we are interested finally only in the on-shell amplitudes,
it suffices to consider the half-on-shell restriction of Eq. (9).
 That is, we put in all the above relations the parameter $z$ onto the initial
energy,
\begin{equation}
z = k^2 / 2\mu - |{\cal E}_0| + i0.
\end{equation}

\noindent
Therefore, the $T^0$-matrix enters Eq. (3) off the energy shell,
differing  thus from the conventional fixed-scatter amplitude
 for which $z = k^2/2\mu$. In scattering-length
calculations we have $k=0$ and hence $z = -|{\cal E}_0| + i0$,
 so that Eqs. (\ref{t0m}) and (\ref{tm3})become nonsingular and easy to
handle. From the practical point of view it is convenient to rewrite Eq.
(\ref{t0m}) by using the Faddeev-type decomposition
\begin{eqnarray}
\nonumber
T^0(\vec{k}',\vec{k};\vec{r};z) &=& \sum^A_{i=1} T^0_i
(\vec{k}',\vec{k};\vec{r};z),
\end{eqnarray}
with \\
\begin{eqnarray}
 T^0_i (\vec{k}',\vec{k};\vec{r};z)
&=& t_i(\vec{k}',\vec{k};\vec{r};z) \
  + \int \frac {d\vec{k}''}{(2\pi)^3}
\frac {t_i(\vec{k}',\vec{k}'';\vec{r};z)} {z - \frac {k''^2}{2\mu}}
\sum_{j\neq i} T^0_j(\vec{k}'',\vec{k};\vec{r};z) .
\label{t0i}
\end{eqnarray}

\noindent
Here, $t_i$ is the t-matrix for the scattering of the $\eta$-meson off the
$i$-th nucleon.\\

As an input information  we need these off-shell amplitudes
and the ground-state wave
functions of the nuclei
involved. Due  to the dominance of the $S_{11}$-resonance near the
threshold energy, we can restrict ourselves to the S-wave
$\eta$N-interaction.
In Ref.\cite{Haid} it was  demonstrated that the role played
by the higher partial
waves is indeed negligible.  The $\eta$NN coupling constant is much
 smaller than the $\eta$NN$^*$ one
\cite{Grei,Alva,Tiat}. This means that the $\eta N$ collision goes
predominantly
 via a virtual formation of N$^*$(1535),\,that is, via the
$\eta$N $\to$ N$^*$ $\to$ $\eta$N reaction.  Hence, the corresponding
amplitude must contain two $\eta$N $\leftrightarrow$ N$^*$
vertex functions and
an N$^*$-propagator in between.  We employ the Yamaguchi-type form
\begin{equation}
t_{\eta N}(k',k;z) = \frac {\lambda}{(k'^2+
\alpha^2)(z - E_0 + i\Gamma/2)(k^2+\alpha^2)}
 \label{tnN}\ ,
\end{equation}

\noindent
commonly used in few-body physics,
with a simple Breit-Wigner propagator.
Two of the four parameters of the t-matrix (\ref{tnN})
are  immediately fixed, namely the resonance energy
$E_0 = 1535$ MeV\,$- \,(M_N + M_{\eta})$ and the width
$\Gamma = 150$ MeV
\cite{PDGr}. The parameter $\lambda$ is chosen   to provide
the correct zero-energy on-shell limit,
i.e., to reproduce the known $\eta N$
scattering length $a_{\eta N}$,

\begin{equation}
 t_{\eta  N}(0,0,0) = - \frac {2\pi}{\mu_{\eta  N}}
 a_{\eta  N}.
\end{equation}
Finally, in order to fix the parameter $\alpha$, we make use of the
results of Refs. \cite{Bhal,Benn}, where the
same $\eta$N $\to$ N$^*$ vertex function ($k^2 + \alpha^2)^{-1}$ was
employed with $\alpha$ being determined via a two-channel fit to the
$\pi $N $\to \pi $N and $\pi $ N $\to \eta $N experimental data. \\

Due to experimental uncertainties and differences
between the models of the
physical processes, there are three different values available
for the scattering length in the literature:
$a_{\eta {\rm N}} = (0.27 + i\ 0.22) fm$ \cite{Bhal},\,\,
$a_{\eta {\rm N}} = (0.28 + i\ 0.19) fm$ \cite{Bhal},
\,\, and  $a_{\eta {\rm N}} = (0.55 + i\ 0.30) fm$ \cite{Wilk}.
For the  range parameter $\alpha$ one also finds  three
different values:  $\alpha = 2.357$ fm$^{-1}$ \cite{Bhal}, \,\,
$\alpha = 3.316$\ fm$^{-1}$ \cite{Benn},\, and $\alpha = 7.617$
\ fm$^{-1} $ \cite{Bhal}.
Since there is no criterium for singling out one of them, we use all 9
combinations of $a_{\eta {\rm N}}$ and $\alpha$
in our calculation.
For the bound states  we employed simple Gaussian - type
functions, which were constructed to be symmetric with respect to nucleon
permutations, and to reproduce the experimental mean square radii:
          $\sqrt{<r_d^2>}$=1.956 fm \cite{rmsD},
          $\sqrt{<r_{{^3H}}^2>}$=1.755 fm \cite{rmsT},
          $\sqrt{<r_{{^3He}}^2>}$=1.959 fm \cite{rmsT}, and
	  $\sqrt{<r_{\alpha}^2>}$=1.671 fm \cite{Til4}.
For masses and binding energies of the nuclei we  used the
experimental values  \cite{Waps}.\\

The method described above involves only the coherent approximation
(7), according to which  all excitations of the
 target nucleus are neglected.
 This approximation could be shown  \cite{Bel5}, as expected, to be
justified at low collision energies and in the case of a wide energy gap
between
the ground and first excited nuclear state.
In our zero-energy calculation the method should, therefore, work
particularly well.
Among the target nuclei $^2$H, $^3$H, $^3$He, and
$^4$He the most accurate results are expected to be  those
for the $^4$He nucleus, since its first excited state at 20.21 MeV
\cite{Til4} lies comparatively high. \\

The results of our  calculations are given in Table I.
As can be seen, the $\eta -$nucleus scattering length depends
strongly on the number of nucleons A, and  is sensitive to the range
parameter $\alpha$, i.e. to the off-shell continuation of the
$\eta N$-amplitude. Prior to this work, Wilkin \cite{Wilk} performed
calculations  of the $\eta $A scattering lengths. In
 these investigations  the $\eta N$ amplitude was simply assumed
 to be constant and equal to the
value of its
zero-energy limit, $a_{\eta N}$=(0.55+i0.30) fm. It is easily  seen
that such an assumption is equivalent to using a zero-range, i.e. a
$\delta$-type interaction. The values of the $\eta\,{^3He}$ and
$\eta\,{^4He}$
scattering lengths obtained by  Wilkin are
$a(\eta\,{^3He}) = (-2.31+ i 2.57)$ fm and $a(\eta\,{^4He}) =
(-2.00+ i 0.97)$
fm.  As our results show, the non-zero-range
character of the
$\eta N$ interaction, i.e., its off-shell properties
change these values significantly.\\

More recently \cite{W94}, Wilkin used the multiple
- scattering expansion in order to incorporate
the nucleon - nucleon correlations. In this approach the
$\eta \alpha$ and $\eta N$  scattering lengths are connected via the simple
relation\\

\begin{center}
$a(\eta,{^4He}) = ( 0.181 / a_{\eta N} - 0.281 fm^{-1} )^{-1}$.\\
\end{center}
\noindent
With the three values of $a_{\eta N}$ listed in Table I, this
formula gives $a(\eta\,{^4He}) =$  ( 0.99 + $i$ 2.67 ) fm,
( 1.39 + $i$ 2.58 ) fm, or (-1.38 + $i$ 6.95 ) fm, respectively.
As can be seen, the agreement between this simple formula and our
microscopic calculation is very poor, especially for large values
of the range-parameter $\alpha$.

If we admit the maximal size of the $\eta N$-interaction region found
in the literature, which corresponds to
$\alpha =2.357$ fm$^{-1}$, and use the rather reasonable
estimate
 $a_{\eta\,N} = (0.55+ i 0.30)$ fm by Wilkins, then the
corresponding large
value of $a(\eta\,{^4He}) = (-4.41+ i 2.86)$ fm
indicates that the condition $A  \geq 12$ obtained in Ref. \cite{Haid}
for the existence of $\eta$-nucleus bound states must be reduced
to lower values of $A$.\\

Since the $\eta N$ interaction is isospin-independent, the difference
    between the $\eta\,^3H$ and $\eta\,^3He$  is
caused by differences of the sizes and binding energies of these nuclei.
 Therefore
 the use of more accurate nuclear wave functions is
of utmost importance. Such realistic wave functions can be
constructed, for instance, by means of the Integrodifferential Equation
Approach (IDEA) \cite{IDEA1,IDEA2}, and are in progress.

\acknowledgments
Financial support by the  University of South Africa (UNISA)
and the Joint Institute  for Nuclear Research, Dubna, is appreciated.
Two of us (S.A.R and W.S) are  grateful to  the Physics
Department of UNISA for  warm hospitality.
\newpage

\newpage

\begin{table}
\begin{tabular}{|c|c|c|c|c|}
\multicolumn{1}{|l|}{} &
\multicolumn{1}{l|} { $\alpha$=2.357 (fm$^{-1}$)} &
\multicolumn{1}{l|} { $\alpha$ =3.316 (fm$^{-1}$)} &
\multicolumn{1}{l|} { $\alpha$ =7.617 (fm$^{-1}$)} &
\multicolumn{1}{c|} { $\alpha_{\eta N}$ (fm)}
\\ \hline
 $ ^2H$   & 0.66+$i$0.82 & 0.65+$i$0.85 &  0.62+$i$0.89   &             \\
 $ ^3H$   & 0.91+$i$1.80 & 0.81+$i$1.88 &  0.63+$i$1.93  &0.27+$i$0.22 \\
$^3He$    & 0.96+$i$1.72 & 0.89+$i$1.78 &  0.76+$i$1.84   &            \\
$^4He$    & 0.90+$i$3.32 & 0.48+$i$3.45 & -0.04+$i$3.40  &            \\
\hline
 $ ^2H$   & 0.75+$i$0.73 &0.74+$i$0.76  & 0.72+$i$0.81    &           \\
 $ ^3H$   & 1.19+$i$1.70 &1.11+$i$1.81  & 0.93+$i$1.92    &0.28+$i$0.19\\
 $^3He$   &1.21+$i$1.59 &1.16+$i$1.68   & 1.04+$i$1.78    &           \\
 $^4He$   & 1.67+$i$3.43&1.23+$i$3.78   & 0.53+$i$3.94   &           \\
\hline
 $ ^2H$    & 1.53+$i$2.00 & 1.38+$i$2.15   & 1.14+$i$2.22    &           \\
 $ ^3H$   & -0.69+$i$5.13 &-1.21+$i$4.50   & -1.30+$i$3.79    &0.55+$i$0.30\\
 $^3He$   &  0.08+$i$5.22 &-0.52+$i$4.83   & -0.79+$i$4.21    &\\
 $^4He$   & -4.41+$i$2.85 &-3.73+$i$2.18   & -3.12+$i$1.85   &\\
\end{tabular}
\end{table}
Table I. The $\eta$-nucleus scattering length results (in fm)
 obtained by using the 9 combinations of the range parameter
$\alpha$ and the $a_{\eta N}$  scattering lenght.


\end{document}